\documentclass[conference]{IEEEtran}
\IEEEoverridecommandlockouts

\usepackage{cite}
\usepackage{amsmath,amssymb,amsfonts}
\usepackage{algorithmic}
\usepackage{graphicx}
\usepackage{booktabs}
\usepackage{textcomp}
\usepackage{xcolor}
\usepackage{url}
\def\BibTeX{{\rm B\kern-.05em{\sc i\kern-.025em b}\kern-.08em
    T\kern-.1667em\lower.7ex\hbox{E}\kern-.125emX}}
\begin{document}

\title{A Lifecycle Cost Analysis of Smart-Contract-Coordinated Federated Learning Marketplaces
\thanks{Identify applicable funding agency here. If none, delete this.}
}

\author{\IEEEauthorblockN{1\textsuperscript{st} Luan Mantegazine}
\IEEEauthorblockA{\textit{Institute of Informatics} \\
\textit{UFRGS}\\
Porto Alegre, Brazil \\
luan.mantegazine@inf.ufrgs.br}
\and
\IEEEauthorblockN{2\textsuperscript{nd} Luiza Leidemer}
\IEEEauthorblockA{\textit{Institute of Informatics} \\
\textit{UFRGS}\\
Porto Alegre, Brazil \\
lsbleidemer@inf.ufrgs.br}
\and
\IEEEauthorblockN{3\textsuperscript{rd} Claudio Geyer}
\IEEEauthorblockA{\textit{Institute of Informatics} \\
\textit{UFRGS}\\
Porto Alegre, Brazil \\
geyer@inf.ufrgs.br}
}

\maketitle

\begin{abstract}

Blockchain-enabled Federated Learning (FL) marketplaces enable collaborative model training among mutually distrustful participants through smart contracts and escrow mechanisms. Although numerous architectures have been proposed for this setting, their economic evaluation is typically limited to the cost of isolated blockchain operations rather than the complete lifecycle of the marketplace. Consequently, it remains unclear whether the operational cost of such platforms depends on operating at scale. This paper presents an experimental study of the operational cost of a Decentralized Autonomous Organization (DAO)-governed marketplace. Our evaluation decomposes the gas consumption of every blockchain operation throughout the contract lifecycle, performs controlled ablation experiments to isolate the impact of the on-chain coordination and the InterPlanetary File System (IPFS) storage layer on federated training, characterizes the scalability of the trainer matching protocol, and derives an analytical model describing the amortization of deployment costs over the federation's operational lifetime. The results show that the lifecycle of a single training task consumes approximately $3.8$ million gas units per hired trainer. The average cost per training round reaches its amortization knee --- defined as twice the asymptotic recurring cost --- after approximately $20$ communication rounds. Moreover, integrating smart contracts and IPFS preserves model performance, achieving a final accuracy comparable to that of a conventional FL deployment without decentralized coordination. These findings demonstrate that smart-contract-coordinated FL marketplaces exhibit an amortizing cost structure not because on-chain operations are inexpensive in absolute terms, but because recurring costs are one to two orders of magnitude smaller than fixed deployment costs and therefore dilute over the operational lifetime of a federation.

\end{abstract}


\section{Introduction}
\label{sec:intro}

Federated Learning (FL)~\cite{1} enables multiple organizations and edge devices to collaboratively train machine learning models without exchanging raw data, making it particularly attractive for privacy-sensitive domains such as healthcare, finance, and the Industrial Internet of Things (IIoT)~\cite{2}. Conventional FL assumes a trusted central coordinator operating within a closed federation. More recently, however, researchers have begun exploring marketplace-based architectures in which independent task requesters and model trainers interact through economic incentives rather than pre-established trust relationships~\cite{3}. In such settings, blockchain platforms provide decentralized coordination, smart contract execution~\cite{4}, and auditable payment mechanisms, enabling collaborative training among mutually distrustful participants~\cite{5}.

Several blockchain-enabled FL marketplaces have proposed mechanisms that address different aspects of decentralized coordination, including hybrid on-chain/off-chain storage based on the InterPlanetary File System (IPFS)~\cite{6,7}, contribution-aware incentive protocols~\cite{2}, escrow- and slashing-based payment policies~\cite{8,9}, and post-quantum trusted execution environments~\cite{10}. Collectively, these studies demonstrate that decentralized FL marketplaces are technically feasible. Nevertheless, existing evaluations primarily focus on isolated architectural components, providing only limited insight into the economic behavior of the complete marketplace lifecycle.

More specifically, prior work typically reports the cost of individual blockchain operations without characterizing the complete lifecycle of a federated learning task. This leads to our central research question --- RQ: is the operational cost of a decentralized FL marketplace dominated by the one-time deployment of smart contracts (and therefore amortizable) or by the recurring operations executed at every communication round (and therefore linear in the number of rounds)? The distinction is fundamental because these two components scale differently with the federation's operational lifetime: a system dominated by fixed deployment costs becomes progressively more cost-effective as training proceeds, whereas one dominated by recurring costs exhibits a nearly linear increase in total cost.

This paper addresses this gap through an end-to-end empirical study of a Decentralized Autonomous Organization (DAO)-governed marketplace, where coordination rules are encoded in immutable smart contracts rather than enforced by a centralized coordinator, IPFS, and the Flower federated learning framework~\cite{11}. Rather than evaluating isolated blockchain transactions, we decompose every blockchain operation executed throughout the marketplace lifecycle, quantify the contribution of each architectural layer through controlled ablation experiments, analyze the scalability of the trainer matching protocol as the participant pool increases, and derive an analytical model that explains how deployment costs are amortized over the federation's operational lifetime.

The main contributions of this work are summarized as follows:

\begin{enumerate}

\item We provide a complete lifecycle decomposition of a smart-contract-coordinated FL marketplace, allowing the operational cost to be separated into one-time deployment costs and recurring execution costs. This decomposition quantifies the contribution of each phase of the marketplace lifecycle to the overall gas consumption.

\item We experimentally isolate the computational overhead introduced by blockchain coordination and IPFS through controlled ablation experiments, demonstrating that both layers have a negligible impact on model convergence while preserving the final prediction accuracy.

\item We derive and validate an analytical amortization model that relates deployment cost to the federation's operational horizon, parameterized by the number of hired trainers ($N_t$). The model demonstrates that the marketplace reaches its amortization knee after approximately $20.37$ communication rounds and that this threshold remains tightly bounded across federation sizes, converging to an upper bound of roughly $22$ rounds.

\end{enumerate}

\section{Related Work}
\label{sec:related}

Research on blockchain-enabled Federated Learning has evolved along three major directions. The first investigates how model artifacts should be stored and referenced in decentralized infrastructures. The second focuses on economic coordination and incentive mechanisms for mutually distrustful participants. The third examines the practical cost of deploying such systems. Collectively, these research directions establish the technical feasibility of decentralized FL marketplaces. However, they evaluate different architectural aspects in isolation, leaving the economic behavior of the complete marketplace lifecycle largely unexplored.

The idea of integrating FL with blockchain was first introduced by Kim \textit{et al.} through BlockFL~\cite{12}, which replaces the centralized aggregation server with a blockchain consensus protocol and established the architectural foundation adopted by subsequent studies. Building on this work, Goh \textit{et al.}~\cite{13} proposed a reference architecture consolidating the hybrid on-chain/off-chain storage model that has since become the dominant design pattern for blockchain-enabled FL.

\subsection{Hybrid storage architectures}

Because neural network parameters are too large to be stored directly on a blockchain, nearly all blockchain-enabled FL systems employ hybrid storage architectures in which model artifacts remain off-chain while the blockchain stores only immutable references~\cite{12,13,14}. Although this architectural pattern is now well established, existing proposals differ in how these references are managed and how participant coordination is performed.

Xia \textit{et al.}~\cite{15} combine the InterPlanetary File System (IPFS) with mutable InterPlanetary Name System (IPNS) references to reduce blockchain interactions while preserving decentralized storage. Their evaluation focuses primarily on communication efficiency and interoperability rather than blockchain execution cost. Ferretti \textit{et al.}~\cite{16} extend this architecture through DeSCo-FL, in which smart contracts dynamically coordinate aggregator election according to predefined scheduling policies. Their work is among the few that report gas consumption, although only for the model registration transaction.

Taken together, these studies establish hybrid on-chain/off-chain storage as the dominant architectural pattern for decentralized FL, yet their evaluations remain transaction-oriented and do not isolate storage cost from coordination cost.

\subsection{Economic coordination and incentive mechanisms}

A second research direction addresses the economic coordination required when task requesters and model trainers do not share prior trust relationships. BlockFL~\cite{12} already introduced on-chain reward mechanisms that compensate trainers according to the amount of contributed data, establishing one of the earliest incentive schemes for blockchain-enabled FL. Since then, escrow-based smart contract mechanisms have become the predominant solution, although existing proposals differ substantially in how participant contributions are evaluated and how payments are released.

FedCoin~\cite{17} replaces conventional blockchain consensus with Proof-of-Shapley (PoSap), using Shapley Value estimation to reward participants according to their contribution to model quality. While the proposed mechanism provides a fair incentive allocation strategy, its specialized consensus protocol prevents direct comparison with standard Ethereum Virtual Machine (EVM) deployments, and its evaluation does not analyze the cost of the complete marketplace lifecycle. FedMarket~\cite{9} generalizes requester--trainer interactions into a competitive pay-as-you-go marketplace, emphasizing market dynamics rather than blockchain overhead. FEDSTR~\cite{18} further extends decentralized AI marketplaces to large language models using the NOSTR ecosystem instead of Ethereum-compatible smart contracts, resulting in a transaction model that differs fundamentally from EVM-based systems.

Other studies improve trust through additional validation mechanisms. FLoBC~\cite{19} introduces dedicated validator nodes that verify submitted models before payment is released, whereas Jaberzadeh \textit{et al.}~\cite{20} integrate FL, IPFS, and smart contracts with slashing policies that penalize dishonest participants. These architectures demonstrate that blockchain can provide computational trust beyond financial settlement. However, their evaluations focus primarily on correctness and security properties, leaving the economic implications of these additional coordination mechanisms largely unexplored.

Overall, these studies rarely distinguish one-time deployment costs from recurring operational costs, preventing a quantitative assessment of how the operational cost evolves over the federation's lifetime.

\subsection{Cost evaluation and scalability}

Cost evaluation remains the least explored research direction in the literature. Existing studies generally report isolated measurements associated with individual smart contract calls or specific architectural components, but rarely characterize the operational cost of the complete lifecycle of decentralized FL systems.

Kalapaaking \textit{et al.}~\cite{21} evaluate smart contract policies for training orchestration, reporting execution time and resource consumption for policy enforcement. Their analysis focuses on governance mechanisms rather than marketplace economics. Wu~\cite{22} presents one of the most comprehensive scalability studies currently available, demonstrating that transferring large models off-chain enables FL to scale to large federations. Nevertheless, the reported cost analysis remains centered on isolated operations rather than the complete contract lifecycle. Coomey and Crosby~\cite{10} investigate the opposite design space by integrating post-quantum cryptography into blockchain-enabled FL. Their evaluation carefully quantifies the gas cost of secure model registration and demonstrates that cryptographic verification contributes only a small fraction of the overall execution time.

More recent studies~\cite{23} follow the same trend by reporting gas consumption for representative operations in scalable and incentive-aware FL frameworks without decomposing the cost of the complete marketplace lifecycle. Collectively, current cost evaluations exhibit three common characteristics. First, gas consumption is typically reported on a per-transaction basis rather than across the entire execution lifecycle. Second, aggregated cost metrics rarely distinguish contract deployment from recurring marketplace operations. Third, comparisons are generally performed across different architectures instead of isolating the contribution of individual architectural layers within the same implementation.

Consequently, despite the substantial progress achieved in blockchain-enabled FL, we found no prior work that explicitly characterizes the extent to which the operational cost of a decentralized FL marketplace depends on scale. This is precisely the RQ stated in Section~\ref{sec:intro}: whether the fixed deployment cost must be amortized over a sufficiently long operational horizon to make the marketplace economically attractive, or whether the platform is already cost-effective from the first rounds due to the intrinsically low cost of recurring operations. Characterizing this amortization behavior is the primary objective of this work.

\section{Architecture}
\label{sec:architecture}

To investigate the operational cost of a smart-contract-coordinated marketplace, we implemented a reference architecture that combines Ethereum-compatible smart contracts~\cite{4}, decentralized storage through the InterPlanetary File System (IPFS)~\cite{6}, and the Flower federated learning framework~\cite{11}. Here, DAO governance denotes the autonomous execution of marketplace coordination rules by smart contracts (participant registration, trainer matching, task creation, and escrow), rather than community-driven governance through token voting. The architecture intentionally follows the design adopted by most blockchain-enabled FL systems, allowing the experimental evaluation to isolate where execution costs are incurred rather than introducing novel coordination mechanisms.

\begin{figure*}[t]
\centering
\includegraphics[width=0.8\textwidth]{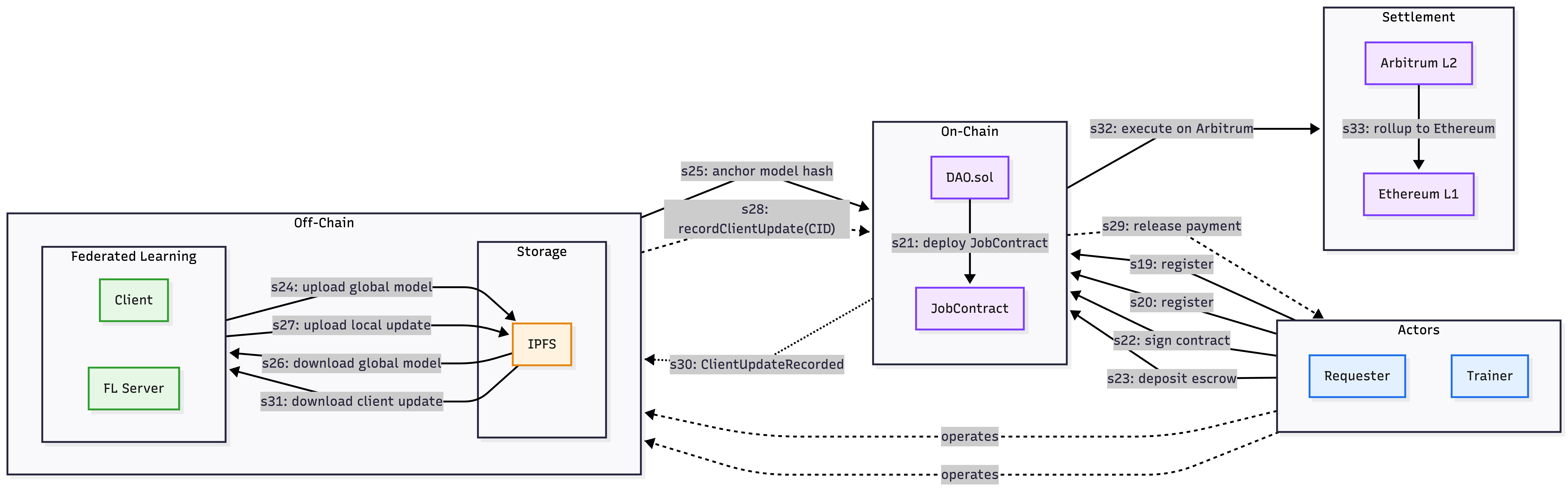}
\caption{Overview of the proposed FL marketplace architecture.}
\label{fig:overview}
\end{figure*}

Figure~\ref{fig:overview} illustrates the operational workflow, organized into three phases with distinct economic characteristics: a setup phase (s19--s23), executed once for each hired trainer; a training loop (s24--s31), repeated at every communication round; and a settlement phase (s32--s33), which operates asynchronously with respect to the other two phases. This phase decomposition anticipates the distinction between fixed and recurring costs quantified in Section~\ref{sec:res-amort}.
 
Setup phase (s19--s23). The requester submits an offer to the \texttt{DAO} contract via \texttt{MakeOffer} (s19). A FL job is instantiated only after a trainer accepts that offer through \texttt{AcceptOffer} (s20), at which point the \texttt{DAO} deploys a dedicated \texttt{JobContract} (s21) and stores its reference internally. Then, requester and trainer complete contract activation by calling \texttt{signJobContract} through the \texttt{DAO}; when called by the requester with funds, this step also locks escrow in the \texttt{JobContract} (s22--s23). After both signatures are recorded, the job reaches the signed state and is ready to coordinate the subsequent FL process.

Training loop (s24--s31). The requester operates the Flower aggregation server, while the trainer executes the corresponding Flower client. At each communication round, the aggregation server publishes the current global model to IPFS (s24) and anchors the corresponding model hash in the \texttt{JobContract} (s25). The trainer retrieves the global model from IPFS through the Flower client (s26), performs local training using its private dataset, and uploads the resulting model update back to IPFS (s27). The trainer then records the corresponding Content Identifier (CID) in the \texttt{JobContract} through the \texttt{recordClientUpdate} function (s28). This transaction automatically releases the corresponding payment to the trainer (s29) and emits the \texttt{ClientUpdateRecorded} event (s30), allowing the aggregation server to identify the newly available model update. The server subsequently downloads the update from IPFS (s31), aggregates it into the global model, and initiates the next communication round.

Settlement phase (s32--s33). Transactions invoking the \texttt{JobContract} are executed on the Arbitrum Layer-2 (L2) network (s32). Periodically, the resulting L2 state is committed to the Ethereum Layer-1 (L1) mainnet through optimistic rollups~\cite{25} (s33), inheriting Ethereum's security guarantees while avoiding the cost of executing every transaction directly on the base layer.

Each execution of steps s19--s23 occurs exactly once for every hired trainer, whereas the training loop (s24--s31) is repeated $R$ times throughout the federated training process, where $R$ denotes the total number of communication rounds. This distinction provides the empirical foundation for the amortization model presented in Section~\ref{sec:res-amort}.

\section{Experimental Setup}
\label{sec:setup}

The experimental infrastructure was designed to enable reproducible and fine-grained measurement of the computational and economic costs associated with the proposed marketplace. Smart contracts were deployed on a local Hardhat~\cite{24} network, which faithfully reproduces the execution semantics of the Ethereum Virtual Machine (EVM) while providing deterministic execution and complete transaction traces. Because the contracts are implemented in standard Solidity, the same bytecode can be deployed without modification on any EVM-compatible blockchain, including Ethereum Layer-2 (L2) rollups such as Arbitrum and Optimism.

Executing the experiments on a local EVM provides two important advantages. First, it enables precise measurement of gas consumption at the opcode level without interference from network congestion or fluctuating transaction fees. Second, it eliminates the monetary cost of repeated experimentation while preserving the execution model of public Ethereum networks. Throughout this work, gas measurements obtained from Hardhat are interpreted as architecture-dependent execution costs. Because the same Solidity bytecode is executed under the standard EVM gas accounting model, the reported gas figures are exact for any EVM-compatible chain rather than approximations. The monetary cost, by contrast, depends on the gas price of the target network; since Layer-2 rollups preserve gas accounting while substantially reducing per-unit fees, the L1 gas price yields a conservative upper bound on the actual monetary cost of deployment.

The decentralized storage layer was implemented using a local IPFS Kubo node. Operating in a controlled storage environment guarantees deterministic availability of stored artifacts and eliminates the variability associated with public IPFS gateways, allowing the experiments to isolate the computational overhead introduced by decentralized storage itself. Table~\ref{tab:components} summarizes the software components used throughout the experimental evaluation.

\begin{table}[h]
\centering
\footnotesize
\caption{Software components used in the experimental evaluation.}
\label{tab:components}
\begin{tabular}{l l}
\toprule
\textbf{Component} & \textbf{Configuration} \\
\midrule
Operating system & macOS (ARM64) \\
Federated learning framework & Flower (Python, gRPC) \\
Local blockchain & Hardhat (EVM) \\
Smart contracts & Solidity, OpenZeppelin \\
Blockchain client & web3.py, eth-account \\
Decentralized storage & IPFS Kubo \\
\bottomrule
\end{tabular}
\end{table}

To isolate the contribution of each architectural layer, the experiments were conducted under three execution configurations.

The \textit{baseline} configuration corresponds to a conventional Flower~\cite{11} implementation of FedAvg~\cite{1}, without blockchain interaction or decentralized storage, thereby incurring no blockchain-related overhead. The \textit{no-IPFS} configuration preserves blockchain-based coordination while replacing IPFS uploads with deterministic content hashes exchanged directly through Flower, thereby isolating the overhead introduced exclusively by smart contract execution. Finally, the \textit{full} configuration implements the complete architecture proposed in this work, in which model parameters are uploaded to IPFS and the corresponding content identifiers are anchored on-chain.

The FL experiments employ the MNIST dataset together with a compact convolutional neural network (MNISTNet). The objective is not to maximize predictive performance but rather to provide a stable and reproducible workload from which communication overhead, blockchain cost, and execution-time variability can be accurately quantified. The training hyperparameters are summarized in Table~\ref{tab:training_params}.

\begin{table}[h]
\centering
\footnotesize
\caption{Federated learning configuration.}
\label{tab:training_params}
\begin{tabular}{l l}
\toprule
\textbf{Parameter} & \textbf{Value} \\
\midrule
Optimizer & Adam ($lr=10^{-3}$) \\
Loss function & NLLLoss \\
Batch size & 32 \\
Local epochs & 1 \\
Training strategies & FedAvg\\
Number of clients & 3 (ablation), 32 (scalability) \\
Communication rounds & 15 \\
Independent runs & 3 random seeds \\
Dataset / Model & MNIST / MNISTNet \\
\bottomrule
\end{tabular}
\end{table}

All experiments were repeated using three independent random seeds. The reported execution times and model accuracies correspond to the arithmetic mean and standard deviation across runs, allowing differences between configurations to be interpreted independently of the stochastic effects introduced by local optimization.

\section{Results and Discussion}
\label{sec:resultados}

This section evaluates the proposed marketplace from four complementary perspectives. First, we investigate the impact of decentralized coordination on the federated training process through a controlled ablation study. Next, we analyze the temporal composition of each communication round and its scalability as the number of participating trainers increases. We then decompose the gas consumption across the complete marketplace lifecycle. Finally, we derive and validate an analytical model that characterizes the amortization of deployment costs over the federation's operational lifetime. Unless otherwise stated, all reported results correspond to the arithmetic mean of three independent executions using different random seeds.

\subsection{Decentralization overhead}
\label{sec:res-ablation}

The ablation study evaluates the individual impact of blockchain coordination and decentralized storage on the federated training process. Three execution configurations are compared: \textit{baseline}, corresponding to a conventional Flower deployment without blockchain or decentralized storage; \textit{no-IPFS}, which preserves blockchain coordination while removing decentralized storage; and \textit{full}, which implements the complete architecture proposed in this work.

Table~\ref{tab:ablation} summarizes the experimental results. All three configurations converge to statistically indistinguishable final accuracies ($0.9882$--$0.9885$), indicating that the proposed coordination mechanisms do not alter the optimization dynamics of the learning process. This observation is consistent with previous studies reporting that blockchain coordination and IPFS preserve the convergence behavior of federated learning~\cite{13}. Such behavior is expected because the aggregation algorithm remains identical across all configurations, with only the communication and coordination mechanisms being modified.

The average execution time per communication round also remains virtually unchanged. The observed differences fall within the experimental variability, indicating that, for compact models such as MNISTNet, the computational cost of local training overwhelmingly dominates the total execution time.

From an economic perspective, only the configurations employing smart contracts incur gas consumption. The small difference observed between the \textit{no-IPFS} and \textit{full} configurations results from dynamic variations in the EVM base fee during the sequential execution of the experiments rather than from any structural architectural difference. For this reason, the subsequent analysis adopts gas consumption per operation as the primary evaluation metric, since it is independent of transient fee fluctuations.

\begin{table}[h]
\centering
\footnotesize
\setlength{\tabcolsep}{4pt}
\caption{Ablation results (MNIST, $N=3$, 15 communication rounds, three independent runs; mean $\pm$ standard deviation).}
\label{tab:ablation}
\begin{tabular}{l c c c}
\toprule
\textbf{Mode} & \textbf{Time/round (s)} & \textbf{Cost/round (ETH)} & \textbf{Final accuracy} \\
\midrule
baseline & $25.05 \pm 1.69$ & $0$ & $0.9882 \pm 0.0005$ \\
no-IPFS  & $24.59 \pm 0.44$ & $4.43\times10^{-6}$ & $0.9884 \pm 0.0006$ \\
full     & $24.77 \pm 0.45$ & $3.73\times10^{-6}$ & $0.9885 \pm 0.0001$ \\
\bottomrule
\end{tabular}
\end{table}

\subsection{Round execution profile and end-to-end scalability}
\label{sec:res-time}

\begin{figure}[h]
\centering
\includegraphics[width=0.55\textwidth]{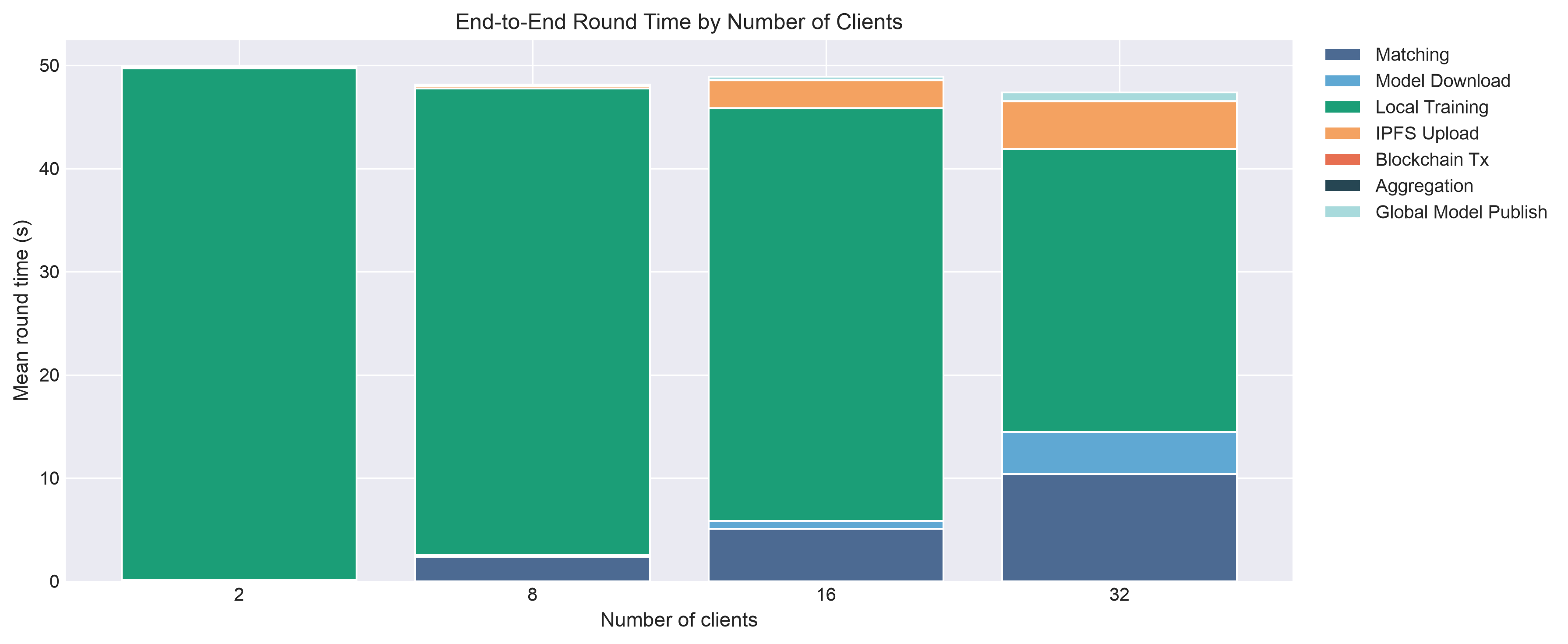}
\caption{Execution-time breakdown as a function of the number of participating trainers ($N\in\{2,8,16,32\}$). Local training dominates the execution time at every scale. The combined off-chain/on-chain overhead remains below $12\%$ even for $N=32$, while the trainer matching stage grows sublinearly and becomes the second largest component for larger federations.}
\label{fig:time-breakdown-scaling}
\end{figure}

Figure~\ref{fig:time-breakdown-scaling} presents the temporal breakdown of a federated communication round for different federation sizes. Regardless of the number of participating trainers, local training remains the dominant component of the execution time. Even for $N=32$, this stage accounts for the largest fraction of the overall execution time, whereas the additional overhead introduced by decentralized coordination represents only a small portion of the total runtime.

As the federation size increases, the time required for trainer matching grows gradually. Nevertheless, the overall execution time per communication round remains nearly constant because local training is performed in parallel across trainers.

These results indicate that decentralized coordination is not the primary performance bottleneck of the proposed marketplace. Instead, local model training continues to dominate the execution profile, while blockchain and decentralized storage introduce only a modest temporal overhead, even for relatively large federations.

\subsection{Marketplace cost decomposition}
\label{sec:res-gas}

\begin{figure*}[t]
\centering
\includegraphics[width=0.95\textwidth]{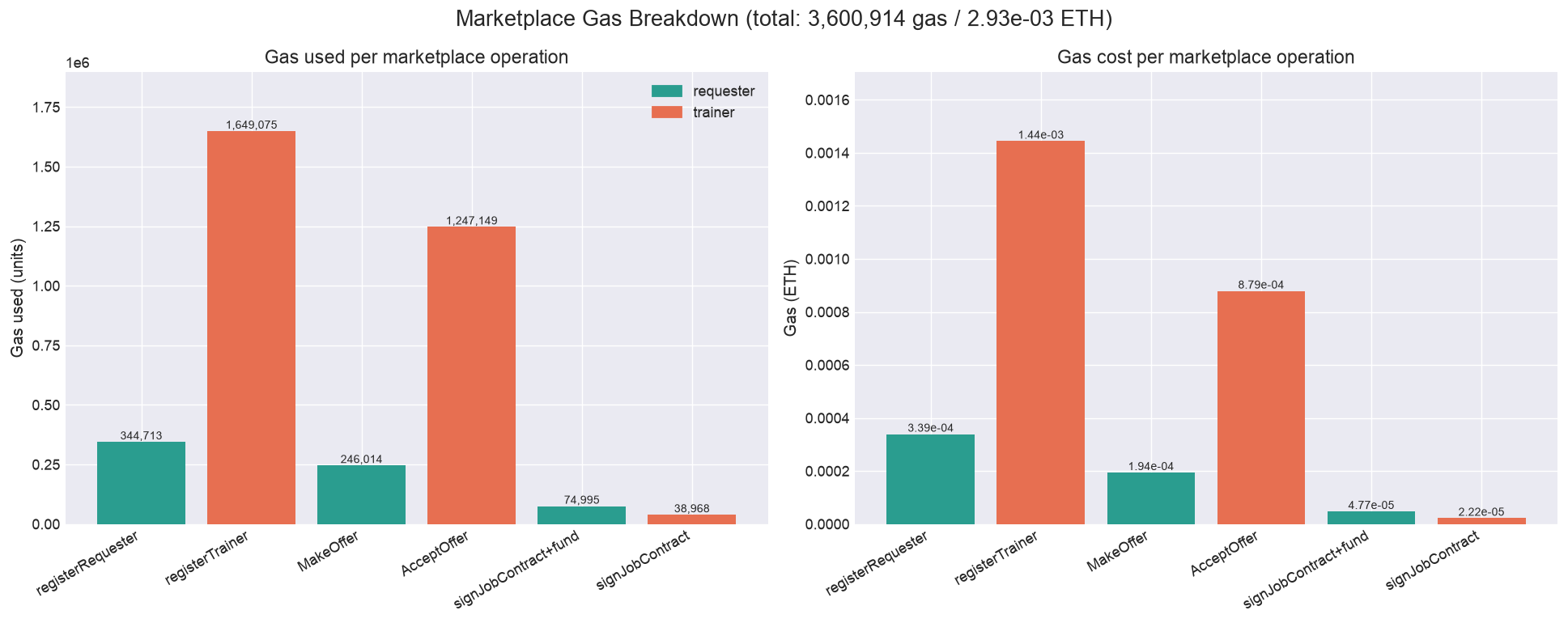}
\caption{Gas consumption of the main marketplace operations, grouped according to the initiating actor (\textit{requester} or \textit{trainer}). Trainer registration and offer acceptance account for most of the deployment cost.}
\label{fig:gas-breakdown}
\end{figure*}

Figure~\ref{fig:gas-breakdown} presents the gas consumption of the main operations executed throughout the marketplace lifecycle. Most of the total gas consumption is concentrated in the initial deployment phase. In particular, the \texttt{registerTrainer} and \texttt{AcceptOffer} transactions account for approximately three quarters of the deployment cost. This behavior results from the creation of persistent storage structures within the Ethereum Virtual Machine, including participant records and task-specific contract metadata.

By contrast, the operations executed during each communication round exhibit substantially lower gas consumption. Recording client updates and publishing the global model primarily modify storage structures that have already been initialized, thereby significantly reducing the gas required by each transaction.

This clear distinction naturally separates deployment costs from recurring operational costs. The former is incurred only once during task creation, whereas the latter remains approximately constant throughout the training process. This observation forms the basis of the amortization analysis presented in the following subsection. It is worth noting that the values reported in Figure~\ref{fig:gas-breakdown} correspond to a single execution of each operation and therefore represent the fundamental cost units used throughout the subsequent economic analysis.

\subsection{Setup cost amortization}
\label{sec:res-amort}

The lifecycle decomposition presented in Section~\ref{sec:res-gas} established the gas consumption of each marketplace operation individually. Building upon these measurements, we derive an analytical model that characterizes how the deployment cost is amortized over the operational lifetime of a federation. We consider the experimental configuration adopted in the ablation study (Section~\ref{sec:res-ablation}), consisting of $N_t=3$ hired trainers, each submitting one model update per communication round.

The fixed deployment cost consists of requester registration, executed only once, together with the per-trainer onboarding cost $G_{\text{setup}}^{*}$, which aggregates trainer registration, offer creation, offer acceptance, contract funding, and contract signing ($G_{\text{setup}}^{*}=G_{\text{regT}}+G_{\text{offer}}+G_{\text{accept}}+G_{\text{sign+fund}}+G_{\text{sign}}$):

\begin{equation}
G_{\text{setup}}(N_t)=
G_{\text{req}} + N_t \times G_{\text{setup}}^{*},
\label{eq:setup-nt}
\end{equation}

\[
G_{\text{setup}}(3)
=
344\,713
+
3\times3\,256\,201
=
10\,113\,316
\]

gas for the experimental configuration.

The recurring cost incurred during each communication round consists of publishing the global model together with recording one client update for every active trainer,

\begin{equation}
G_{\text{round}}(N_t)
=
G_{\text{publish}}
+
N_t
G_{\text{record}},
\label{eq:round-nt}
\end{equation}

resulting in

\[
G_{\text{round}}(3)
=
54\,533
+
3\times147\,289
=
496\,400
\]

gas per communication round.

This decomposition naturally leads to the average gas consumption per communication round over a federation lasting $R$ rounds,

\begin{equation}
\bar{g}(R)
=
\frac{G_{\text{setup}}(3)}{R}
+
G_{\text{round}}(3)
=
\frac{10\,113\,316}{R}
+
496\,400,
\label{eq:amort}
\end{equation}

which explicitly separates the fixed deployment cost from the recurring operational cost.

\begin{figure}[h]
\centering
\includegraphics[width=0.5\textwidth]{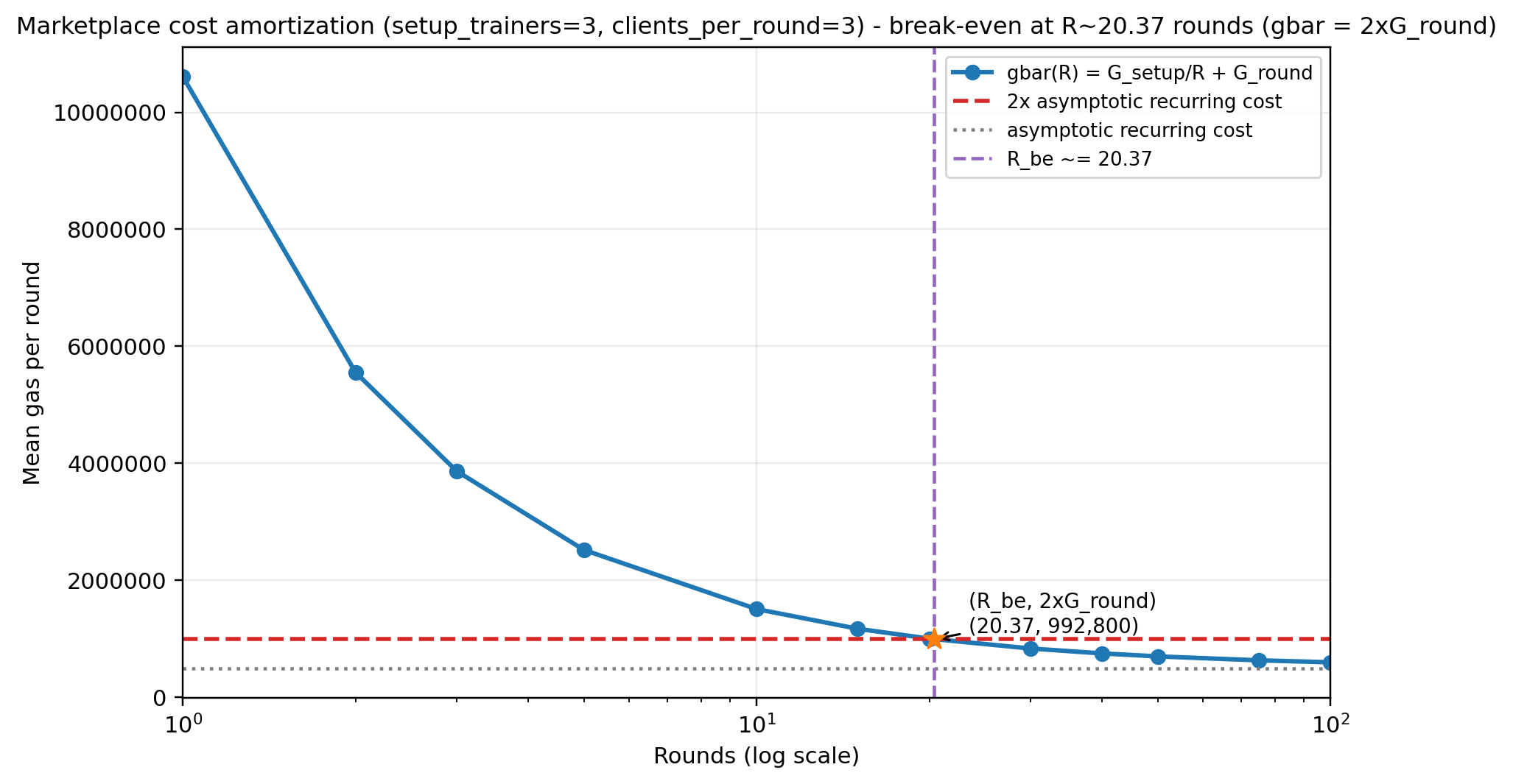}
\caption{Marketplace cost amortization for $N_t=3$ trainers and three updates per communication round. The left panel shows $\overline{g}(R)=G_{\text{setup}}/R+G_{\text{round}}$ on a logarithmic scale with respect to $R$, together with the reference lines $G_{\text{round}}$ (asymptote) and $2\,G_{\text{round}}$ (knee threshold), as well as the geometric indication of $R_{be}\approx 20.37$. The right panel displays only the cumulative gas consumption, isolated to avoid the misleading impression of an intersection between the two curves.}
\label{fig:amort}
\end{figure}

Table~\ref{tab:amort} evaluates Eq.~\eqref{eq:amort} for representative operational horizons, while Figure~\ref{fig:amort} illustrates the corresponding continuous amortization curve together with the amortization knee.

\begin{table}[h]
\centering
\footnotesize
\setlength{\tabcolsep}{4pt}
\caption{Setup cost amortization as $R$ increases for $N_t=3$ hired trainers. The setup fraction is defined as the ratio $G_{\text{setup}}(3) / G_{\text{total}}(R)$, while the ratio $\overline{g}/G_{\text{round}}(3)$ quantifies how much the average cost per communication round exceeds the asymptotic recurring cost.}
\label{tab:amort}
\begin{tabular}{r r r r r}
\toprule
$\mathbf{R}$ & \textbf{Gas total} & $\overline{\mathbf{g}}(R)$ & \textbf{Setup frac.} & $\overline{\mathbf{g}} / G_{\text{round}}$ \\
\midrule
5   & $12{,}60$M & $2{,}52$M & $80{,}3\%$ & $5{,}07\times$ \\
10  & $15{,}08$M & $1{,}51$M & $67{,}1\%$ & $3{,}04\times$ \\
20  & $20{,}04$M & $1{,}00$M & $50{,}5\%$ & $2{,}02\times$ \\
50  & $34{,}93$M & $699$k    & $29{,}0\%$ & $1{,}41\times$ \\
100 & $59{,}75$M & $598$k    & $16{,}9\%$ & $1{,}20\times$ \\
\bottomrule
\end{tabular}
\end{table}
 
\begin{figure}[h]
\centering
\includegraphics[width=0.48\textwidth]{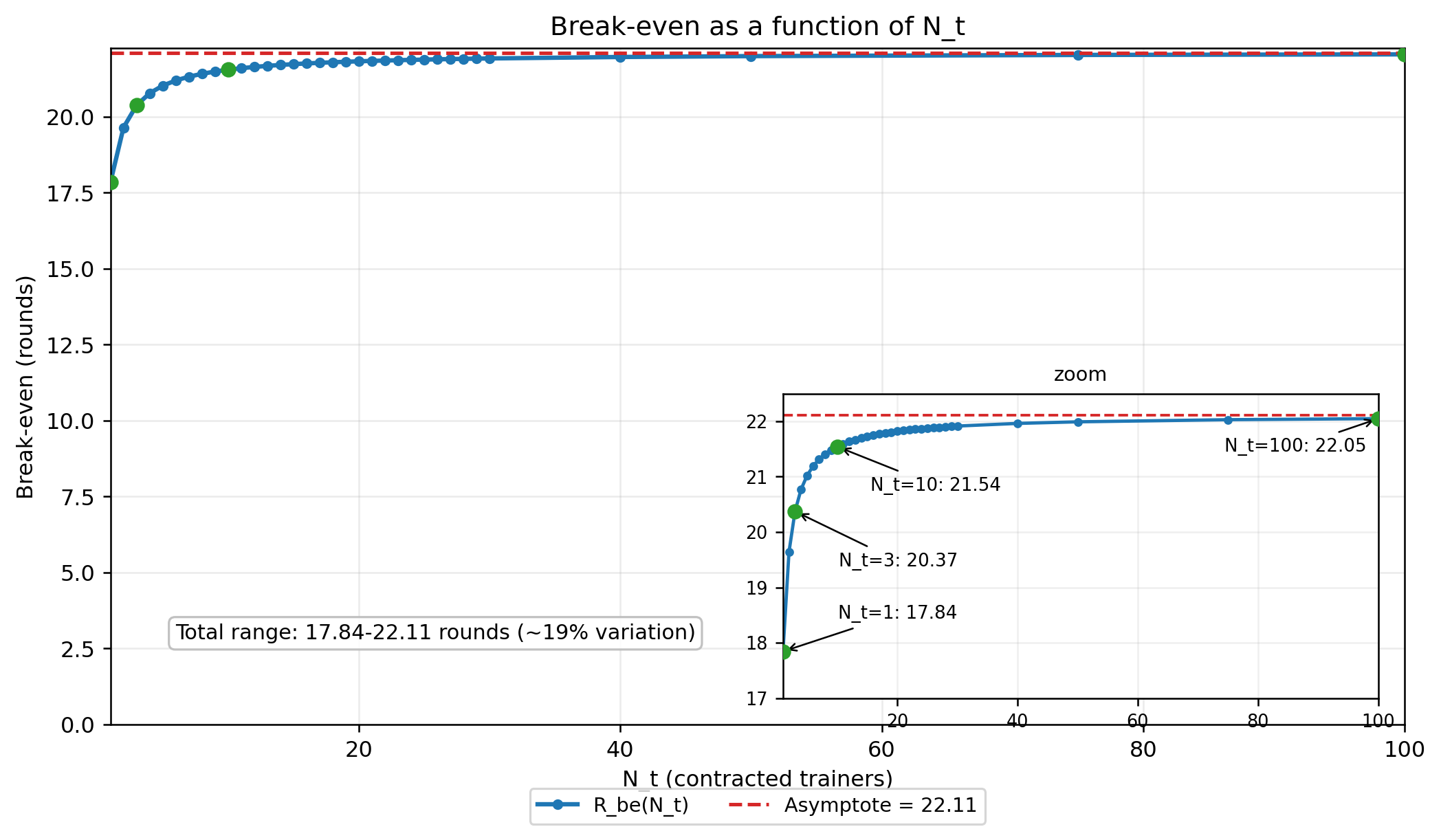}
\caption{Amortization knee parameterized by the number of hired trainers ($N_t$), with the main axis starting at zero and an inset (\textit{zoom}) provided for improved readability. Under the measured operating regime ($\bar{k}=1$ pending offer per trainer upon acceptance), the knee values are $R_{be}(1)=17.84$, $R_{be}(3)=20.37$, $R_{be}(10)=21.54$, and $R_{be}(100)=22.05$, converging to the asymptote $\lim_{N_t\to\infty} R_{be}(N_t)=22.11$---a variation of only $\sim\!19\%$ across two orders of magnitude in $N_t$.}
\label{fig:breakeven-nt}
\end{figure}
Three observations emerge from these results.

First, the average cost per communication round decreases rapidly as the federation operates for additional rounds. At $R=5$, the average cost is still $5.07\times$ larger than the asymptotic recurring cost, whereas at $R=100$ this ratio decreases to only $1.20\times$.

Second, the marketplace exhibits a well-defined amortization knee,

\[
R_{be}
=
\frac{G_{\text{setup}}(3)}
{G_{\text{round}}(3)}
\approx
20.37
\]

communication rounds, at which the average cost per round equals twice the asymptotic recurring cost. We adopt the $2\times$ threshold because at that horizon the fixed deployment cost contributes exactly as much to the per-round average as the entire recurring cost, marking a natural boundary between the deployment-dominated and recurrence-dominated regimes. Beyond this operating horizon, the relative contribution of deployment rapidly diminishes, accounting for less than $30\%$ of the total cost at $R=50$ and less than $17\%$ at $R=100$.

Third, the deployment component asymptotically approaches zero as the federation lifetime increases. Consequently, sufficiently long-running marketplaces operate at essentially the recurring cost alone.

\paragraph{Knee robustness with respect to federation size.}

Because the previous analysis considers the specific experimental configuration ($N_t=3$), we generalize the amortization knee as a function of the number of hired trainers by combining Eqs.~\eqref{eq:setup-nt} and~\eqref{eq:round-nt},

\begin{equation}
R_{be}(N_t)
=
\frac{
G_{\text{setup}}(N_t)
}{
G_{\text{round}}(N_t)
}
=
\frac{
G_{\text{req}}
+
N_tG_{\text{setup}}^{*}
}{
G_{\text{publish}}
+
N_tG_{\text{record}}
},
\label{eq:rbe-nt}
\end{equation}

where $G_{\text{setup}}^{*}=3\,256\,201$ gas denotes the trainer onboarding cost.

Figure~\ref{fig:breakeven-nt} presents the resulting knee horizon for $N_t\in\{1,\ldots,100\}$.

The analysis reveals that the knee horizon remains tightly
bounded as the federation size increases. Across two orders of
magnitude in the number of hired trainers, the amortization knee
varies only from $17.84$ to $22.05$ communication rounds,
converging to the upper bound

\begin{equation}
\lim_{N_t\rightarrow\infty} R_{be}(N_t)
=
\frac{G_{\text{setup}}^{*}}{G_{\text{record}}}
\approx
22.11,
\label{eq:rbe-limit}
\end{equation}

as follows directly from Eq.~\eqref{eq:rbe-nt}: both deployment and
recurring costs grow linearly with the number of hired trainers,
causing their ratio to rapidly approach a constant value.
Consequently, the amortization behavior is primarily determined by
the marketplace architecture rather than by federation size itself.
This result holds under the measured operating regime, in which each
trainer has, on average, one pending offer at the time of acceptance
($\bar{k}=1$).

\paragraph{Threats to validity.}
Four aspects bound the generality of these results. First, the evaluation uses a single compact model (MNISTNet) on MNIST; because $G_{\text{publish}}$ and $G_{\text{record}}$ depend on the size of the anchored artifact reference rather than on the model itself, the per-round cost is expected to remain stable across model sizes, but this was not varied experimentally and the absolute deployment figures should not be read as model-independent. Second, MNIST was chosen to provide a stable and reproducible workload rather than to maximize predictive performance (Section~\ref{sec:setup}), so the accuracy-preservation result speaks to the coordination layer, not to task difficulty. Third, the amortization analysis assumes the measured operating regime of one pending offer per trainer at acceptance ($\bar{k}=1$); Eq.~\eqref{eq:rbe-nt} indicates that the knee horizon is sensitive to this parameter, and other matching regimes were not explored.

More importantly, this analysis directly answers the research question raised in Section~\ref{sec:related}. The operational cost of Layer-2 FL marketplaces does not stem from intrinsically inexpensive blockchain operations. On the contrary, the deployment cost is substantial. Instead, operational cost arises because this fixed investment is paid only once for each hired trainer and is progressively amortized over the federation lifetime. Consequently, the economic sustainability of smart-contract-coordinated FL marketplaces is determined primarily by the structure of their cost model rather than by the absolute cost of individual blockchain transactions.

\section{Conclusion}
\label{sec:conclusao}
 
This work evaluated the operational cost of a Decentralized Autonomous Organization (DAO)-governed federated learning (FL) marketplace on an Ethereum Layer-2 network. Unlike prior studies reporting the cost of isolated blockchain operations, we decomposed the complete marketplace lifecycle into deployment and recurring execution phases. The cost proved overwhelmingly concentrated in deployment: onboarding a single hired trainer requires approximately $3.8$ million gas units, mostly to create persistent EVM state during registration and job contract deployment, whereas recurring per-round operations are substantially cheaper, causing the average cost per round to approach its recurring asymptote after approximately twenty communication rounds. The decentralization mechanisms did not interfere with learning: smart contracts and IPFS preserved the final accuracy with only a modest time overhead, and local optimization remained the dominant component of each round.
 
Taken together, these findings show that the operational cost of decentralized FL marketplaces depends primarily on cost amortization rather than on intrinsically inexpensive blockchain transactions. The deployment cost is significant but incurred only once per hired trainer and progressively diluted over the federation lifetime. Lifecycle-level cost decomposition and amortization analysis therefore provide a more informative basis for evaluating decentralized coordination architectures than aggregate transaction costs alone.

\end{document}